# Electroluminescence as a probe of strong exciton-plasmon coupling in few-layer WSe$_2$

*Yunxuan Zhu*[1,†], *Jiawei Yang*[1,†], *Jaime Abad-Arredondo*[2,†], *Antonio I. Fernández-Domínguez*[2], *Francisco J. Garcia-Vidal*[2], *Douglas Natelson*[1,*]

[1]Department of Physics and Astronomy, Rice University, Houston, TX 77005, USA.

[2]Departamento de Física Teórica de la Materia Condensada and Condensed Matter Physics Center (IFIMAC), Universidad Autónoma de Madrid, E-28049 Madrid, Spain.

[†]These authors contributed equally to this work.

[*]Corresponding author: Douglas Natelson (natelson@rice.edu).

**ABSTRACT**

The manipulation of coupled quantum excitations is of fundamental importance in realizing novel photonic and optoelectronic devices. We use electroluminescence to probe plasmon-exciton coupling in hybrid structures consisting of a nanoscale plasmonic tunnel junction and few-layer two-dimensional transition-metal dichalcogenide transferred onto the junction. The resulting hybrid states act as a novel dielectric environment that affects the radiative recombination of hot carriers in the plasmonic nanostructure. We determine the plexcitonic spectrum from the electroluminescence and find Rabi splittings exceeding 50 meV in strong coupling regime. Our experimental findings are supported by electromagnetic simulations that enable us to explore systematically, and in detail, the emergence of plexciton polaritons as well as the polarization characteristics of their far-field emission. Electroluminescence modulated by plexciton coupling provides potential applications for engineering compact photonic devices with tunable optical and electrical properties.

It is well known that two oscillators can couple together to generate new normal modes through the exchange of energy. The same analogy applies to quantum emitters, or emitter ensembles, when placed inside an optical cavity, sustaining photonic or plasmonic modes[1]. In the weak coupling regime, their spontaneous emission rate is modified by the density of electromagnetic (EM) states inside the cavity[2,3]. From optical interactions with the emitter(s)[4,5], the cavity spectrum may develop a dip at the excitonic frequency (exciton-induced transparency) which becomes more apparent in the strong coupling regime[6–8]. Polaritonic mixed states emerge that are part light, part matter, and the cavity spectrum undergoes Rabi splitting into two separate polariton branches. The optical response of the strongly coupled system is highly sensitive to the state of the photon emitter(s)[7], which provides a means of manipulating quantum states of light and can enable high-fidelity quantum operations [9–11] and nonclassical photon generation[12,13].

Transition metal dichalcogenides (TMDs) are ideal materials for coupling to optical resonators, as TMDs interact strongly with light through sharp excitonic modes with high binding energies of a few hundred meV even at room temperature[14], particularly in the monolayer limit[15–18]. Embedding the strong dipole moment of these optical transitions in EM resonators has enabled the formation of hybrid, polariton states whose matter component are TMD excitons[19]. Accessing individual strongly coupled emitters has been limited largely to far-field optical probes under diffraction-limited light beams[20–23], which impedes miniaturization and large scale integration for practical applications. Electrically driven plasmonic nanostructures are a feasible route to realize and probe nanoscale strong coupling phenomena. By placing exciton-supporting ultrathin TMDs in close proximity to localized surface plasmons (LSPs), it is possible to access the LSPs near field, opening new possibilities for the tuning, study, and functionality of strongly coupled nanoscale systems. Near-field electrical probes are a promising tool to characterize polaritonic physics in TMD platforms, as conventional far-field optical methods can pick up additional emission from uncoupled excitons.

Here, through current-driven hot carrier electroluminescence (EL) of a plasmonic junction[24–26], we access the strong coupling regime in the interaction between the LSPs[27] of the junction and the TMD excitons. The sub-nm sized gap between the two electrodes serves as an ultra-confined plasmonic nanocavity where incoherent photons are generated by hot carrier recombination. When the TMD is coupled in the near field to the nanogap LSPs, the resulting plexciton polaritons strongly modify the radiative local density of states that governs light emission from the hot carriers in the driven plasmonic metal. The EL emission process thus acts as an extremely local, near-field probe of plasmon-exciton polariton physics and a new means of controlling the flow of energy at the nanoscale through changing the dielectric environment. Measured spectra are consistent with a theoretical model parameterized through EM simulations describing the coupling of LSPs and TMD excitons, building the connection between the nanoscale hot carrier dynamics in the metal and the photons radiated into the far field. These findings open new paths towards compact solid-state quantum light sources and strong coupling-based control of hot carriers in plasmonic nanostructures.

A 3D sketch of the setup and a brief schematic showing the coupling mechanics are illustrated in Fig.1a (detailed experimental methods and measurement description can be found in section 5 in SI). An example EL spectrum with unpolarized detection, obtained at 0.8-0.9 V applied bias from a bare tunnel gap *without* $WSe_2$, is shown in Fig. 1b. At low currents the photon emission from a biased nanogap is generally attributed to the radiative decay of surface plasmon modes excited inelastically by tunneling electrons. In the low current limit, individual tunneling electrons should only excite plasmon modes up to a cutoff energy given by the applied bias $eV$, leading to the "below-threshold" light emission ($\hbar\omega \leq eV$)[28–31]. Continuum emission at energies higher than $eV$ implies that another mechanism also contributes to the far-field emission: the plasmon-enhanced radiative recombination of hot carriers (electrons and holes) in the vicinity of the tunnel junction in the high current limit[24–26,29,31–33]. In this mechanism, the EL spectrum embeds the rich plasmon mode structure of the system. The broad peak that appears in the bare junction at around 1.69 eV is assigned to a LSP resonance associated with transverse dipolar optically bright mode of the metal nanowire with well-defined width (~120 nm)[24,34]. While there is device-to-device variation in the overall plasmonic mode structure, this dipolar LSP appears consistently in this energy range (set by nanowire width) in these nanogap structures[34]. Longitudinal across-gap dipolar plasmon modes at lower photon energies, 1.2 – 1.5 eV, are also apparent in the spectra. These are energetically far from the exciton, as discussed below, and do not affect the exciton-plasmon coupling near 1.69 eV.

The photoluminescence (unpolarized detection) from a bare trilayer $WSe_2$, excited by a 532 nm laser, is also plotted in Fig. 1b. The sharp peak appearing at 1.68 eV corresponds to the direct A exciton transition energy, consistent with previous studies[35–37]. The spectral overlap between the broadened transverse-mode EL peak in the bare plasmonic junction and the PL exciton in the plain TMD opens the way towards the emergence of strong light-matter coupling phenomena in the hybrid system.

An image of the TMD-coupled plasmonic junction structure is shown in Fig. 1c. An optical microscope image of the hybrid structure immediately after dry transfer for $WSe_2$ is shown in Fig. S1 in Supporting Information (SI). The thickness of the transferred $WSe_2$ flake is characterized using the atomic force microscopy (AFM), as can be seen in Fig. 1d. The height profile shown in the lower part of Fig. 1d reveals a thickness of ~3.9 nm for trilayer $WSe_2$, also consistent with previous studies[37].

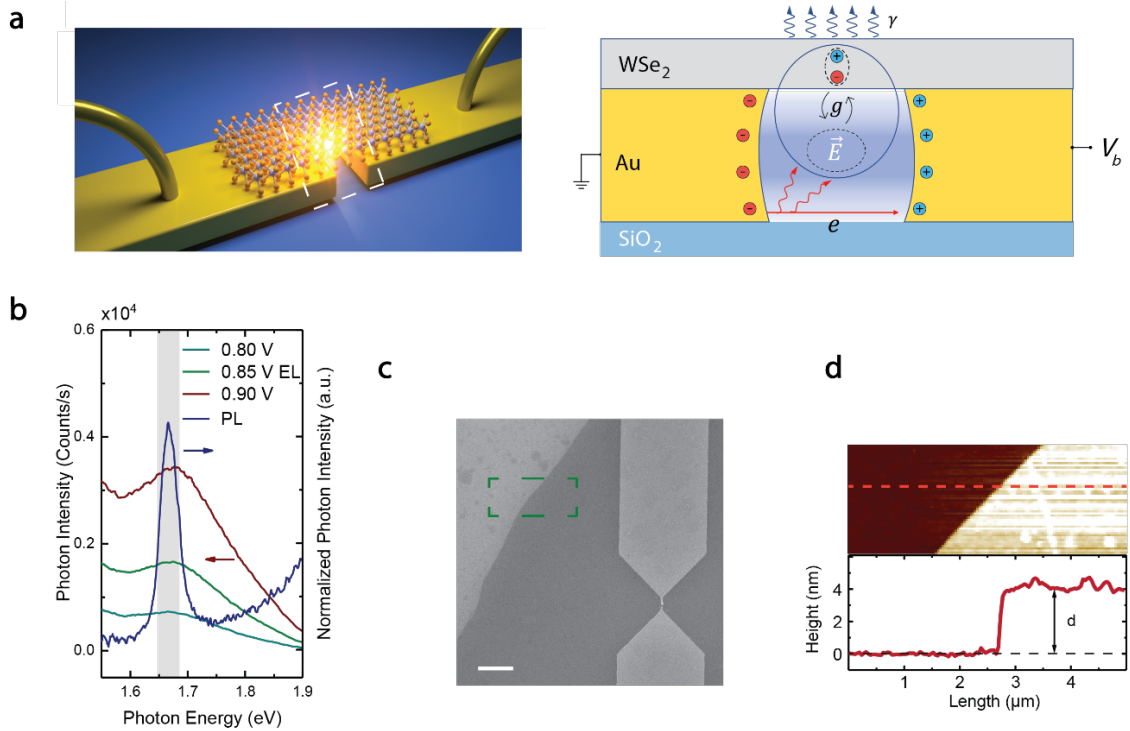

**Figure 1.** Schematics of experimental setup and various characterization for the TMD and EL spectrum. (a) schematic of a fabricated hybrid TMD-on-gap structure and the diagram of plasmon-exciton coupling connected to electrically driven tunneling within the gap. (b) EL spectra (no polarization selection) for a bare gold junction at different biases with a zero-bias conductance of 0.11 $G_0$, plotted together with the PL spectrum (no polarization selection) for a plain bilayer of $WSe_2$. Shaded area indicates the full width at half maximum of the A exciton peak. (c) Electron microscopy image of an electromigrated junction with TMD on top. The scale bar in the figure is 2 μm. (d) AFM image of the transferred TMD edge indicated by the green dotted lines in (c). Below shows the cross-section height profile of the red dashed line.

When the coupled TMD-nanogap structure is biased to the EL regime, instead of the single broad plasmonic resonance in this energy range routinely seen in bare metal junctions, we observe two peaks in the emission (no polarization selection) for this device, reproducing the Rabi splitting phenomenology of strongly coupled systems (Fig. 2a). These two peaks emerge around 1.69 eV, associated with two plexcitonic-like states that result from the coupling between the transverse dipolar LSP mode of the junction and the excitons in the $WSe_2$ flake. This device shows a spectral separation of 50 meV, which allows us to estimate a collective plasmon-exciton coupling strength of around 25 meV[1]. Doublet spectra for different bias voltages are shown together with the PL spectrum for the A exciton of this flake of $WSe_2$. We have also used $MoS_2$ as an alternative for WSe$_2$ to form a TMD-gap system with a blue-tuned plasmonic resonance (narrower nanowire width) to accommodate the higher exciton energy (~1.9 eV). Analogous polaritonic spectra can be observed, as shown in Fig. S2 in the SI. Note that the bias voltages are far below the energy scale of the EL emission, demonstrating that this is the hot carrier regime, in which photons are generated by recombination of hot carriers in the metal.

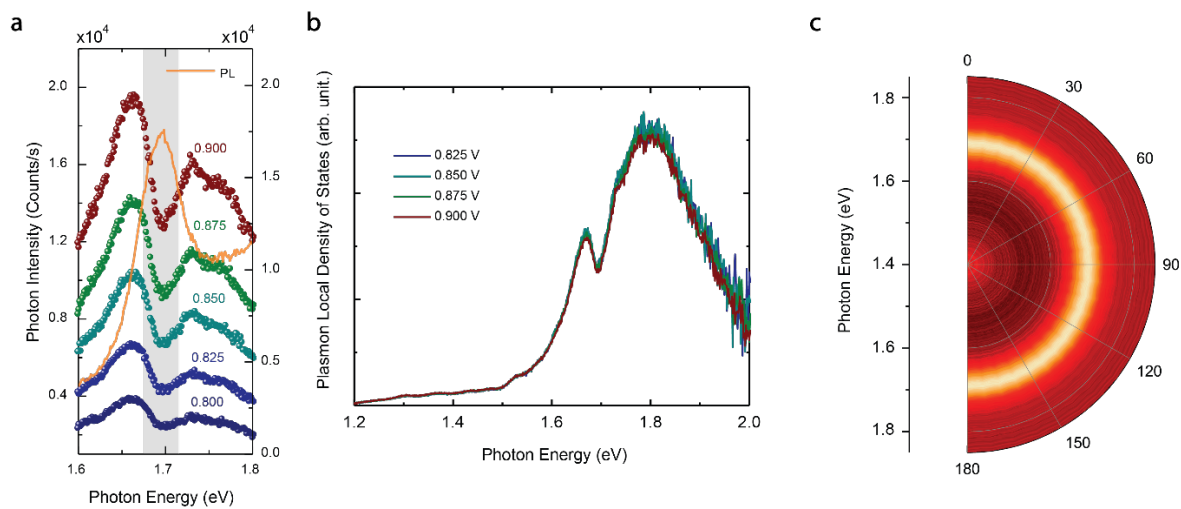

**Figure 2**: Experimental EL spectra, inferred plasmonic density of states, and PL spectra. (a) Measured EL spectra (no polarization selection) at different biases from 0.8 V to 0.9 V with a zero-bias junction conductance of 0.20 $G_0$, plotted together with the PL spectrum measured for the WSe2 on top. (b) Extracted optical radiative local density of states obtained by removing the Boltzmann-like hot carrier energy distribution for different biases. (c) Normalized polarization dependent contour map of PL spectra at the nanogap obtained by changing the detecting polarization from 0 to 180 degrees. Laser's polarization is fixed parallel to the nanogap. The isotropic exciton PL peak at ~1.7 eV does not show any signature of LSP coupling nor an appreciable polarization dependence.

To confirm how the plasmon-exciton coupling is manifested through hot carrier EL, we have performed normalization analysis to extract the radiative local density of states $\rho(\omega)$ of the hybrid TMD on the gap system based on previous work[24,31] (the detailed process is described in SI section 1 and Fig. S3). As can be seen in Fig. 2b, the extracted $\rho(\omega)$ in the polariton energy range all collapse for different biases. This consistency validates that the emission spectrum scales as the product of a polariton-modified $\rho(\omega)$ and a voltage-dependent hot carrier Boltzmann distribution in the metal. This confirms that the radiative decay of electrically generated hot carriers in the metal can be manipulated by introducing excitonic states in the near field of the plasmonic nanostructure. This mechanism of plasmonic hot carrier electroluminescence coupled into a polariton-modified $\rho(\omega)$ is distinct from and complementary to the generation of exciton luminescence via plasmon-exciton resonant energy transfer[38]. The plexcitonic impact on the photonic density of states is analogous to effects seen in other examples of plasmon-exciton coupling[39–41]. The plasmonic nanogap by itself has a photonic density of states with sharp features because of the nanogap geometry and confinement of the nanogap LSP modes. When coupled to the TMD exciton, the photonic density of states of the hybrid system is modified via the effective permittivity of the local environment, resulting in split polaritonic modes. Hot carriers are generated through the non-radiative damping of the electrically excited surface plasmons when the electrons inelastically tunnel through the nanogap. Radiative photons produced via radiative recombination within the steady state hot carrier distribution are emitted into the plexciton-shaped local photonic density of states of the hybrid TMD/nanogap system through the Purcell effect. Therefore, it is possible to control the hot carrier energetic decay path in plasmonic nanostructures through strong plasmon-exciton coupling.

For better comparison, polarization-resolved PL spectra are taken at the nanogap (Fig. 2c). This far-field PL is fully governed by emission from uncoupled TMD excitons, far from the nanogap junction but within the incident beam spot. The polarization-resolved PL is obtained by changing detection polarization through a linear polarizer in front of the spectrometer while fixing excitation laser's polarization at 0 degree (parallel to the nanogap). The isotropic exciton peak intensity in PL spectra with no splitting feature across all the polarization angles indicates that the far-field-excited PL shows no clear signs of the nanoscale plasmon-exciton coupling environment.

The EL from the hybrid TMD-nanogap system's small mode volume gives access to extremely local information that is inaccessible via far-field methods.

Numerical simulations (described below) show that the system behind the spectral features in Fig. 2a enters the plasmon-exciton strong coupling regime and forms plexciton polaritons. This conclusion is further supported by the EL measurements presented in Fig. S4 in SI, which correspond to samples with WSe$_2$ flakes of different thicknesses. These reveal that the doublet profile only develops for few-layer (less than 5) WSe$_2$ samples, while exciton absorption would become stronger with increasing TMD thickness.

To gain physical insight into the observed plasmon-exciton coupling phenomenology, we have performed numerical simulations in COMSOL Multiphysics. (detailed modeling can be found in the SI). Real and imaginary parts of the in- and out-of-plane WSe$_2$ permittivity are shown in Fig. 3a. Fig. 3b plots the emission spectrum obtained from the numerical simulations (solid line). We observe that the numerical results reproduce the experimental features in Fig. 2a. Most importantly, in agreement with the experiments, $I_{ff}(\omega)$ presents an asymmetric doublet-like feature, with an emission dip at 1.7 eV.

We further employ our EM model to perform a systematic analysis of the plasmon-exciton coupling strength to identify in which light-matter coupling regime (weak or strong) our device is operating. Fig. 3c and d presents two different studies that investigate whether the strong-coupling regime is reached and well-developed plexciton polariton states are formed in the experimental samples. Figure 3c presents the usual anti-crossing map broadly employed in the literature to identify the onset of strong coupling. In our calculations, we swept the WSe$_2$ exciton frequency from 1.4 to 2 eV, keeping the rest of the model parameters. Black dotted lines plot both frequencies. We fitted the numerical emission map to the expression of the intensity spectrum obtained for a simplified model consisting of a single excitonic emitter interacting with a single plasmonic resonance[6]. As a result of the fitting, we extracted a plasmon-exciton coupling strength $g = 31$ meV, consistent with the experimental values.

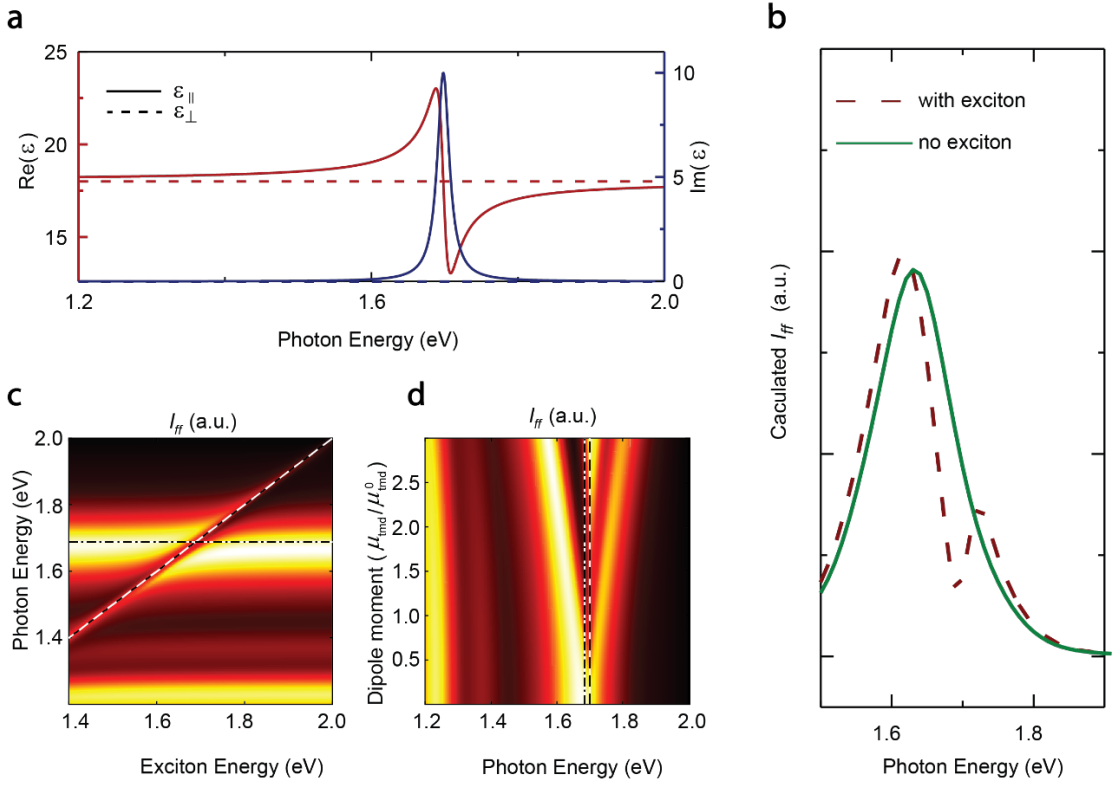

**Figure 3**: Numerical models of emission spectra for the hybrid TMD-on-gap structure. (a) WSe2 permittivity (real and imaginary parts) as utilized in our EM model. (b) Numerical emission spectrum, $I_{ff}(\omega)$, for the model TMD-plasmonic system. Solid (dashed) line corresponds to calculations including (excluding) the excitonic contribution to the WSe2 permittivity. (c-d) Numerical results for the emission $I_{ff}$ maps by sweeping the model exciton frequency (c), and dipole moment (d). The colors code the far-field intensity from minimum (black) to maximum (yellow) in a linear, thermal color scale.

Fig. 3d presents a similar study, but now sweeping the exciton dipole moment. The value taken in the TMD permittivity in Fig. 3a is labelled as $\mu_{tmd}^0$. For $\mu_{tmd} \ll \mu_{tmd}^0$, the spectrum presents a single peak at 1.7 eV, a dip emerges with increasing dipole moment, initially due to exciton absorption and subsequently, because of polariton formation. For $\mu_{tmd} \gg \mu_{tmd}^0$, two polariton branches are clearly apparent, and another emission peak emerges in between the Rabi doublet.

This feature can be attributed to the light scattered by WSe$_2$ excitons that remain uncoupled to the plasmon resonance, described through a strong variation in the real part of $\varepsilon_{\parallel}(\omega)$ in our model. Such a small peak feature at the exciton energy can also be seen in measured polarized spectra shown in Fig. S7, which are line cuts of the polar contour data plots shown in Fig. 4b. The fitting of this map yields the same plasmon-exciton coupling strength as the previous one at $\mu_{tmd} = \mu_{tmd}^0$. The surface plasmon linewidth extracted from both maps is $\gamma_{SP} = 90$ meV, which allows us to conclude that the experimental samples are in the plasmon-exciton strong coupling regime, based on the strong coupling criterion $4g > \gamma_{SP} + \gamma_{tmd}$[8,42].

Polarization-selected EL spectra facilitate resolving the spectral fingerprint of LSP-exciton strong coupling in the presence of other plasmon modes. The electromigration process yields atomic scale variations (protrusions and bumps) within the nanogap, causing device-to-device variations in the relevant plasmon mode's linewidth in EL spectra and rich spectral mode features at different detected polarizations[24,34]. Polaritonic splitting features only appear in spectra obtained from devices incorporating the TMD; full polarization-selected spectra of 20 bare-metal junctions show no polariton-like peak splittings (see Fig. S8). Polarization-dependent spectral measurements can thus provide further insights into the exciton coupling with the LSP modes.

The polarization dependence of the EL spectra can be well reproduced through a two-plasmon model, which is the interaction between a TMD exciton and a LSP mode, together with emission by a single, weakly confined surface plasmon polariton mode. Notice that only the LSPs can be treated accurately in our EM calculations as the simulation volume ranged only up to 2 microns. The guided surface plasmon mode is blue-tuned and decoupled from the TMD exciton (See SI section 3). It is generated at the tunnelling junction and propagates away from it via the gold nanowire and electrodes. Note that the propagation lengths of surface plasmons in gold stripes of dimensions comparable to our samples is of the order of 10 microns[43]. As explained above, by construction, the far-field emission of this guided mode cannot be described by our EM calculations. The polarization of its emission into the far-field is expected to vary from device to device, as it results from lateral beating effects with phase characteristics fixed in the mode excitation at the atomic-scale features of the tunnelling junction. As the bowtie-like electrode geometry widens along the propagation direction away from the junction, more and more propagating modes with different symmetries along its width are sustained (with different numbers of nodes and antinodes along the transverse direction). The leaky plasmonic waves then involve the beating between these modes with different transverse symmetries, which would radiate with different polarization characteristics. The polarization of emission from the non-coupled guided mode would then depend very much also on how the nanowire and electrode widths varies along the propagation direction, apart from the initial phase profile they acquire in the junction region. To deal with the different length scales in the experiments, we built an illustrative model based on a master equation formalism that accounts for all radiative losses (details of the derivation can be found in SI Sect. 6.2).

As shown in Fig. 4, the polarization-selected EL spectra of three representative TMD-coupled devices show clear signatures of polaritonic splitting that can be reproduced using the two-plasmon model. The EM simulations successfully reproduce the low-energy (around 1.25 eV) LSP mode and the TMD-coupled LSP resonance at 1.63 eV (see Fig. S9 and S10). The non-coupled guided plasmon mode in all three devices is assumed to be at 1.79 eV with a linewidth $\gamma_{gp} = 140$ meV, and the dominant polarization dependence of its emission is allowed to vary from device to device (90 degrees in panel a, 130 degrees in panel b, and 180 degrees in panel c), in agreement with simulations studying the bare nanowire behavior (Fig. S11 and S12). Figure S9 shows a detailed comparison of the EL spectrum at different detected emission polarizations with the two-plasmon model for the device in Fig. 4a. Relative peak heights of the upper and lower plexciton contributions change as polarization angle evolves. Polarization selection in detection reveals the onset of strong coupling in the system by filtering out far-field contributions from the guided modes that are not interacting with the TMD excitons. We note that the polaritonic splitting can be considerably larger than the exciton PL linewidth, which is clearly shown by the linecuts of Fig. 4(a-c) presented in Fig. S14. In Sect. 4 of the SI, a discussion on the role of exciton absorption and its distinction from plexcitonic spectral splitting is provided.

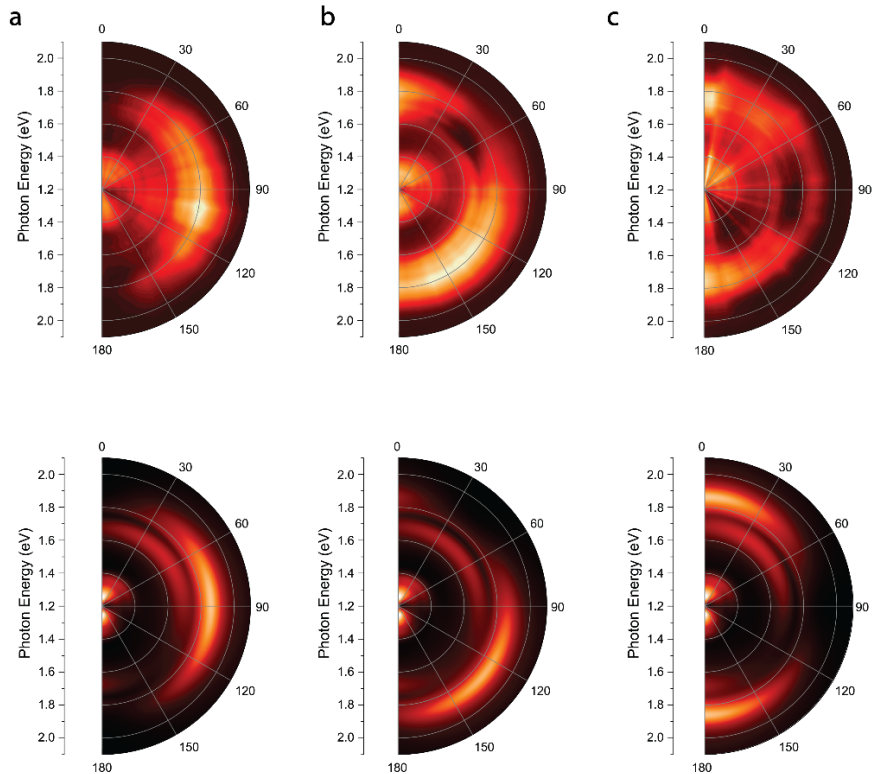

**Figure 4**. Unnormalized full contour plot of the experimental results and the numerical calculated polarization-resolved far field intensity. (a-c) the upper panel shows the polar plot for the experimental measured spectra, while the lower panel shows the numerical results obtained from the theoretical formalism. The strong coupling feature is polarization dependent and is most visible at different polarization angles for three devices (~70 degrees in panel a, 45 degrees in panel b, and 90 degrees in panel c), which can be explained by the variance of the guided surface plasmon mode's dominant polarization (90 degrees in panel a, 130 degrees in panel b, and 180 degrees in panel c).

Hot carrier recombination in EL plasmonic nanogaps is shaped by the photonic density of states for emission. In this work, we show that optical interactions between such a nanogap and an adjacent TMD layer can reach the strong coupling regime, and that the EL acts as a new near-field probe to access such extreme local information, with plexciton formation tuning the recombination of hot electrons and holes in the metal. Far-field PL measurements on these devices are not sensitive to the coupling of the extremely localized nanogap modes to the TMD excitons. The richness of the experimental plasmonic spectrum shapes the polarization dependence of the far-field plexcitonic EL spectrum. While in the present design there is device-to-device variation in the polarization dependence of the EL, all TMD-coupled devices can be modeled with a simple two-plasmon approach, with a LSP resonance coupled to the TMD excitons, and an uncoupled guided plasmon mode at higher energy with dominant polarization that is allowed to vary with device. Polarization-selected spectra reveal plexcitonic spectral features by filtering out the device-specific guided mode. Further pushing the coupling strength should be possible if the design is changed to sandwich the TMD in between the nanogap electrodes, a nanogap analogy to the nanoparticle-on-TMD-on-mirror geometry. These results open avenues for fabricating novel on-chip electroluminescent heterostructures that leverage and control plasmon-exciton coupling through proper nanoscale geometric engineering. Plexcitonic effects, by determining the local photonic density of states, can be designed to manipulate the energetic relaxation of hot carriers in the metals supporting surface plasmons.

**ASSOCIATED CONTENT**

**Supporting Information**.

The following file is available free of charge on the ACS Publication website. Hot carrier induced electroluminescence; Control experiment and theoretical modelling for hBN spacer; Theoretical insights into the blue-detuned mode; Discussion regarding the mode splitting: plexcitonic strong coupling vs excitonic absorption; Experimental methods; Theoretical modelling and computational techniques.

## AUTHOR INFORMATION

### Corresponding Author

* corresponding author: Douglas Natelson Email: natelson@rice.edu

### Author Contributions

D.N., J.Y. and Y.Z. designed the experiment. J.Y. and Y.Z. fabricated the devices, conducted the experiment, and modelled the data. F.J.G.-V., A.I.F.-D. and J.A-A. theoretically analyzed and numerically modeled the experimental polaritonic light emission spectra and polarized coupling spectra. All authors wrote the manuscript and have given approval to the final version of the manuscript.

### Data Availability Statement

The data underlying this study are openly available in the Zenodo archive at https://doi.org/10.5281/zenodo.10371451.

### Funding Sources

ONR N00014-21-1-2062, Robert A. Welch Foundation Award C-1636, NSF ECCS-2309941. This work has been also funded by the Spanish Ministry of Science, Innovation and Universities (AEI) through grants PID2021-126964OB-I00 and PID2021-125894NB-I00, by Comunidad de Madrid through the Proyecto Sinérgico CAM 2020 Y2020/TCS-6545, and by the ``(MAD2D-CM)-UAM7'' project funded by the Comunidad de Madrid, by the Recovery, Transformation and Resilience Plan from Spain, and by NextGenerationEU from the European Union. J.A.-A. acknowledges funding from the Spanish MU (FPU18/05912 scholarship).

**Notes**

The authors declare no competing financial interest.

**ACKNOWLEDGMENT**

D.N. J.Y. and Y.Z. acknowledge ONR N00014-21-1-2062; D.N. and Y. Z. acknowledge Robert A. Welch Foundation Award C-1636 and NSF ECCS-2309941. J.A.-A. acknowledges funding from the Spanish MECD (FPU18/05912 scholarship).

**ABBREVIATIONS**

LSPs, localized surface plasmons. EM, electromagnetic. TMDs, transition metal dichalcogenides. EL, electroluminescence. PL, photoluminescence. hBN, hexagonal boron nitride.

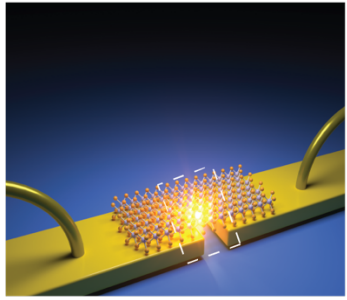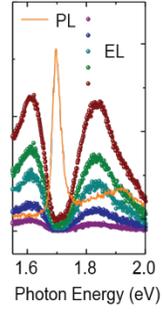

TOC figure

# Supporting Information

# Electroluminescence as a probe of strong exciton-plasmon coupling in few-layer WSe$_2$

*Yunxuan Zhu[1,†], Jiawei Yang[1,†], Jaime Abad-Arredondo[2,†], Antonio I. Fernández-Domínguez[2], Francisco J. Garcia-Vidal[2], Douglas Natelson[1,*]*

[1]Department of Physics and Astronomy, Rice University, Houston, TX 77005, USA.

[2]Departamento de Física Teórica de la Materia Condensada and Condensed Matter Physics Center (IFIMAC), Universidad Autónoma de Madrid, E-28049 Madrid, Spain.

[†]These authors contributed equally to this work.

[*]Corresponding author: Douglas Natelson (natelson@rice.edu.)

**Supplementary Information Text**

**1. Hot carrier induced electroluminescence**

There are multiple mechanisms that can produce electroluminescence from a biased plasmonic tunnel junction[1–6]. Inelastic tunneling electrons excite localized surface plasmons (LSP) localized in the tunneling gap. The excited LSP energy is limited by the applied voltage bias ($E_{ine} < eV_b$). The LSP can then go through either the radiative decay channel to form far field radiative photons[3], or non-radiative damping to generate hot electrons and holes away from the Fermi level[1]. These hot carriers subsequently recombine and generate photon emission on a timescale of tens of femtoseconds, determined by the hot carrier relaxation. The flow of energy through the electronic system depends on the time interval between successive tunneling electrons (magnitude of the tunneling current density) and plasmonic strength of the junction[2]. In the present case, we purposely apply electrical excitations to drive the system into the hot carrier emission regime (i.e., high tunneling current and strong plasmonic performance), where the generated hot carriers form a steady-state population of nonequilibrium in both source and drain electrodes. Radiative recombination from those hot carriers produces a continuum spectrum with an energy that extends well above the applied bias ($eV_b$), of the form expressed by Eq. (2).

In this limit, the steady-state distribution of those hot carriers can be expressed by a Boltzmann distribution, with the effective temperature defined in the Boltzmann factor ($\exp(-\hbar\omega/k_b T)$) proportional to applied electrical bias. It has been demonstrated clearly in our previous work that one can perform a normalization analysis to separate the contribution of the voltage-dependent Boltzmann factor and voltage-independent plasmonic local density of states $\rho(\omega)$ in Eq. (2). By dividing the measured spectrum at different biases with a reference spectrum acquired at one particular bias voltage yields

$$ln\left(\frac{I_{ff}(\omega)}{I_{ff,ref}(\omega)}\right) = -\frac{\hbar\omega}{k_B}\left(\frac{1}{T_{eff}} - \frac{1}{T_{eff,ref}}\right) \qquad (S.1)$$

where $I_{ff,ref}(\omega)$ and $T_{eff,ref}$ corresponds to the reference spectrum. After extracting the effective temperature from Eq. (S.1), $\rho(\omega)$ can also be obtained through Eq. (2) in the main text.

As can be seen in Fig. S5b, the normalized spectra decays linearly on a logarithmic scale versus the photon energy, with extracted effective temperature labeled adjacent to each line. The extracted optical radiative local density of states based on the effective temperature collapsed to a single voltage-independent spectral for all the biases across the whole spectral energy range (Fig. 2b). This indicates the plasmon exciton coupling is manifested through effectively modifying local photonic density of states, which is subsequently embedded in the far field radiation spectrum mediated by the hot carriers. It should be noted here that Fig. 2b actually plots $\omega^4 \rho(\omega)$ as it absorbs the pre-factor written in Eq. (2).

## 2. Control experiment and theoretical modelling for hBN spacer.

To investigate the plasmon-exciton coupling further in our samples, control experiments were performed in which hexagonal boron nitride (hBN) layers were introduced as a spacer between the nanogap and the WSe$_2$. As shown in the sketches in Fig. S5, exfoliated hBN with two thicknesses were encapsulated in between the TMD and the gold nanowire using the dry transfer procedure described in Methods. The dip disappears when hBN thickness increases, verifying that the plasmon modes sustained by the gold nanostructure are strongly localized at the junction and present a decay length along the vertical direction of only a few nanometers. This is also confirmed by the numerical results shown in Fig. S6, which are consistent with the experimental results in Fig. S5, predicting the vanishing of plexcitonic spectral signatures for a sufficiently thick dielectric spacer between TMD and metal nanowire.

We can employ COMSOL simulations to investigate the effect that introducing a dielectric spacer between the gold nanostructure and the WSe$_2$ flakes has in the emission spectrum, in the same spirit as the experimental implementation of this shown in Fig. S4. To avoid other possible effects, the hBN spacer is modelled as a non-excitonic layer (a non-excitonic version of the TMD layer in Fig. 3), with $\varepsilon_{\parallel}(\omega) = \varepsilon_{\perp}(\omega) = 18$. The thickness of the WSe$_2$ is fixed to 5 nm, and different layers of this artificial material is placed between the TMD and the gold junction. Figure S6 shows the calculated impact of the introduction of 1 and 2 nm thick spacers in the far-field intensity. For clarity, $I_{ff}(\omega)$ is plotted in log scale. The Rabi-like doublet apparent in the sample with no spacer is strongly suppressed and deformed with the introduction of only a 1 nm thick (~2 layer) hBN spacer. For a 2 nm spacer (~4 layer), the splitting profile has disappeared completely, and an emission peak (of very low intensity) emerges at the exciton frequency, which again, can be related to uncoupled TMD excitons. These results are in qualitative agreement with the experimental observations, in which the emission dip presented a very small contrast for a hBN thickness between 1 and 2 nm.

## 3. Theoretical insights into the blue-detuned mode

To investigate the origin of the blue-detuned mode (~1.85 eV) that is not appreciably coupled to the exciton but mainly contributes to the polarization behavior, we have performed a set of numerical simulations to explore the optical modes sustained by the gold nanowire in the absence of the nanogap, as a function of nanowire (or strip) width and with a fixed thickness of 18 nm. For that, we remove the nanogap from our simulation geometry described in the methods section and set up a plane wave excitation (polarized normally to the nanowire axis). By measuring the scattered power, we can identify the relevant radiative modes of the strip. Fig. S11 shows the scattered power as a function of photon energy and nanowire width. Note that in these simulations, the incident field polarization is matched to the resonances that are not present in our near-field simulations. To identify the main contributions, the maps are fitted to a four-mode Lorentzian decomposition, and the resulting frequencies are shown in black solid lines. For narrow stripes (<100 nm), the maximum scattering takes place at ~1.6 eV. For wider nanowires similar to the experimental samples (>150 nm), the scattering maximum is at ~1.3 eV. We can identify this with the prominent peak apparent in the experimental spectra in Fig. 2 and the calculations in Fig. 3. These structures support higher frequency modes, which yield a less pronounced scattering maximum at ~1.8 eV. We believe that these are responsible for the high frequency peaks in Fig. 4a. Fig S12 shows the near-field modal electric field maps for different wire widths. These show their guided, propagating character, presenting multiple lobes along the nanowire edges. We can also observe how the low-frequency modes in narrow stripes are able to accommodate to the bow-tie geometry of the structure, the lobes are apparent beyond the nanowire section. On the contrary, the guided modes scatter (and radiate) at the nanowire extremes at higher frequencies in wider structures, and the lobes cannot be observed in the near-field. Moreover, the varying thickness of the electrodes, imperfections and roughness in the experimental samples increase the radiative losses experienced by these modes.

**4. Discussion regarding the mode splitting: plexcitonic strong coupling vs excitonic absorption**

In this section, we provide a detailed discussion on the observation of spectral splitting in the main text, and we eliminate the possibility that the splitting results from excitonic absorption rather than plexcitonic coupling. To unambiguously explain that our observed splitting is not due to absorption, we need to show that the magnitude of the polaritonic splitting can vary from less than to greater than the exciton PL linewidth. If the mode splitting originated from excitonic absorption, the splitting linewidth should mirror the exciton linewidth in the PL spectra. The exciton PL linewidth is either much larger or narrower compared to the mode splitting for all three devices (Fig.4 a, b and c), as can be seen in Fig. S14. The full width at half maximum (FWHM) of the exciton PL peak and the mode splitting are denoted as '$a$' and '$b$', respectively, with the numerical relationship displayed on the top left corner of the lower panels ($b \sim 0.9a$ for (a), $b \sim 2.6a$ for (b), and $b \sim 2.9a$ for (c)). This confirms that the linewidth of the PL is not in general the same magnitude as the mode splitting, and the polaritonic splitting can vary from less than the exciton PL linewidth to more than twice the exciton PL linewidth. Therefore, the experimental data demonstrate that the observed mode splitting is not due to absorption, and the system clearly enters in the strong-coupling regime on when the splitting is much larger than the PL linewidth.

Due to the difficulty in altering the nanogap geometry continuously in the atomic scale once it is created through electromigration (which is different from previous studies[7]), it is hard to show the continuous evolution of the experimental splitting from zero to a value larger than the exciton linewidth in one device. However, emission spectra from various devices can have various splitting linewidths ranging from narrow to broad due to the way that TMD is coupled microscopically to the nanogap and the detailed atomical geometry inside the gap We further note that linecuts at different detected polarization angles (Fig. S9) illustrate how the splitting's visibility can evolve as a function of a particular parameter (polarization angle) due to the coupling of the exciton to particular plasmonic modes, as explained in our model. This kind of polarization dependence of the splitting's visibility has been seen in other exciton/plasmon systems[8] at or near strong coupling.

For additional evidence that the splitting in emission for the TMD/nanogap hybrid system does not originate from exciton absorption, we have also performed control experiments for the hybrid system with various thickness of $WSe_2$, as can be seen in Fig. S4. $WSe_2$ with various thicknesses are transferred on top of the nanowires for the electroluminescence measurements, and the doublet emission profile only develops for few-layer (less than 5) WSe$_2$ samples, and completely vanishes with increased WSe$_2$ thickness. This is consistent with our theoretical prediction that emission in this system is not dominated by excitonic absorption, because increasing the WSe$_2$ thickness would significantly enhance absorption.

## 5. Experimental methods

### 5.1 Device fabrication

The fabrication of the nanowire devices starts with a Si wafer that has 300 nm thick thermal oxide on top. The nanowires were first patterned using standard e-beam lithography. Briefly, double layer PMMA (495/950 combination) e-beam resist was applied to provide an undercut cross-section profile after development to ensure clean lift off the pure gold (18 nm thick) after evaporation (without any adhesion layer to improve the plasmonic performance and maximize the local density of photonic states[1,9].). The wire dimensions are pre-determined so that the nanostructure supports plasmon modes that are resonant in the energy range of the TMD excitons (for WSe$_2$ at ~ 1.7 eV, 18 nm thick, 120 nm wide, 600 nm long). Each nanowire is connected to bow tie-shaped fan outs with long extended electrodes (150 μm), to facilitate wire bonding and the few-layer WSe$_2$ dry transfer process. WSe$_2$ flakes with various thicknesses are mechanically exfoliated onto a polydimethylpolysiloxane (PDMS) stamp and located using an optical microscope. WSe$_2$-coupled nanowires are then formed by employing a dry transfer method to drop down the selected WSe$_2$ on top of the nanowire to create the final device for strong-coupling measurements. The TMD on top of the nanogap will crack when the electrical bias across the gap exceeds 1.6 V during the measurement of EL spectrum possibly due to increased Joule heating, as can be seen in Fig. S1 of the supplemental information. Therefore, all the EL measurements are performed below 1.2V bias to keep the whole TMD-nanogap hybrid system intact.

**5.2 Nanogap formation**

The fabricated arrays of nanowire samples were mounted into a high vacuum ($\sim 10^{-9}$ mbar) optical cryostat. Initially at a substrate temperature of 80 K and then again at 30 K, the electromigration technique[1] was applied to form a sub nanometer sized gap within the nanowire device. Specifically, cycles of voltage (0 to 1 V sweep) were applied across the nanowire and halted when a sharp drop (>5%) recorded in the current flowing in the nanowire. Subsequently, the same process was repeated until the conductance drops slightly below 1 $G_0 = 2e^2/h$ (to ensure the formation of the tunnel junction), followed by the EL and PL measurements of the prepared device.

**5.3 Optical measurements**

A 100× long working distance objective with NA of 0.7 was used to collect photon emission from the biased TMD-nanowire devices. The emitted photons were guided through free space optics based on a home-built Raman spectroscopy setup[10] and focused onto a Si CCD spectrometer (Horiba iHR 320) for data collection. The spectra shown in all the figures are already corrected for the energy-dependent quantum efficiency of the CCD. PL spectra of the TMD was excited by a CW 532 nm laser through the collecting objective, and the map scan was performed through a lens mounted on a piezo, so that the focal spot was scanned with tens of nanometer precision.

**6. Theoretical modelling and computational techniques**

**6.1 Numerical simulations**

To gain insight into the electromagnetic resonances supported by the junction and how these couple to the TMD excitons, we have performed numerical simulations in the Maxwell's Equation solver in COMSOL Multiphysics. Our EM model mimics the geometrical dimensions of the experimental samples with the metal permittivity taken from previous work[11]. The nanojunction is composed by a 600 nm long and 120 nm wide gold strip. It is connected at the edges to electronic contacts that branch out at 45 degrees. To create the nanojunction, we cut the strip at an angle of 14º and create a gap of 14 nm. This gap is created at a lateral offset of 190 nm from the center of the strip to break the symmetries, as in the experimental samples. We then add two half cylinders of 5 nm radius to each of the faces of the cut to create a pico-cavity with a minimum gap distance between the cylinders of 4 nm. The metallic structure has a height of 18 nm. To simulate the tunneling current, we place a dipolar source at the center of the 4 nm gap and set the dipole moment pointing from one face of the gap to the other. The metallic structure material is set to gold described by the permittivity given by[11]. The gold structure is set over a substrate made of $SiO_2$ described by a refractive index of 1.5. On top of the metallic junction, we place a 5 nm layer of $WSe_2$, described by the permittivity given in the main text. On top of the $WSe_2$ we have air ($n = 1$). The total simulation domain is a sphere of 1 $\mu$m radius centered around the source dipole position and is terminated by a scattering boundary condition. To obtain the detected intensity we calculate the radiative Purcell, $P_R(\omega)$, or the local density of radiative states, obtained by integrating the time-averaged Poynting vector along the vertical (top) direction. Optical properties of $WSe_2$ are modelled through a Clausius–Mossotti anisotropic dielectric function[12] that assumes that the exciton dipolar moments are oriented in-plane of the TMD layer. Thus, we have in- and out-of-plane permittivity components of the form $\varepsilon_{||}(\omega) = \varepsilon_b \frac{1+2\beta(\omega)}{1-\beta(\omega)}$ and $\varepsilon_\perp(\omega) = \varepsilon_b$, respectively, with

$$\beta(\omega) = \frac{\mu_{tmd}^2}{3\varepsilon_0 \hbar} \rho_{tmd} \frac{2\omega_{tmd}^2}{\omega_{tmd}^2 - (\omega + i\gamma_{tmd}/2)^2} \qquad (1)$$

where the different constants were obtained from the fitting to the experimental data in Ref. [13]. Thus, $\varepsilon_b = 18$ is the background permittivity, and the exciton transition frequency, $\omega_{tmd}$, and linewidth, $\gamma_{tmd}$, were set to 1.7 eV and 20 meV, respectively, in agreement with the bare PL spectrum in Fig. 1a. Real and imaginary parts of the in- and out-of-plane WSe$_2$ permittivity are shown in Fig. 3a. Notice that such a small value for $\gamma_{tmd}$ implies that the excitonic contribution to the real and imaginary parts of $\varepsilon_\parallel(\omega)$ is relevant only within a spectral band of 20 meV. The weight of the exciton dipole moment and exciton density is given by $\mu_{tmd}^2 \rho_{tmd} = 10^{-4}$ e$^2$/nm. These values are in agreement with recent studies[14].

Adapting the phenomenological expression obtained from previous experiments on hot carrier EL[1,2], we define the numerical far-field emission intensity as

$$I_{ff}(\omega) = \omega^4 \rho_{rad}(\omega) \exp\left(-\frac{\hbar \omega}{k_B T_{eff}}\right) \qquad (2)$$

where $\rho_{rad}(\omega)$ is the local density of radiative photonic states at the electron tunnelling light source, obtained from our numerical calculations of the radiative Purcell factor, and $T_{eff} = 1400\ K$[1], in agreement with the experimental estimates.

### 6.2 Two mode coupling model

To fully capture the far-field experimental phenomenology we have developed a model. This is meant to reproduce experimental features that, according to our numerical simulations, are not related to the strong light-matter interactions that take place within the micron-sized volume around the nanowire junction. In particular, the model allows us to introduce a far-field maximum observed in the experiment, blue-detuned from the TMD exciton frequency, that we link to the strong radiation losses experienced by the surface plasmon waves that scatter at the nanowire edges, microns away from the region of electron tunnelling. By construction, this maximum is absent in the numerical simulations, where the lateral boundary conditions prevent any power radiation in the vertical direction originated from propagating surface plasmons.

In our model, we have included two optical modes, and the exciton transition. The Hamiltonian that describes the interaction and pumping of the system is given by, $H = H_0 + H_{pump}$, with

$$H_0 = \hbar\omega_1 a_1^\dagger a_1 + \hbar\omega_\sigma \sigma^\dagger \sigma + \hbar g\left(a_1^\dagger \sigma + a_1 \sigma^\dagger\right) + \hbar\omega_2 a_2^\dagger a_2,$$

$$H_{pump} = \hbar v_1\left(a_1^\dagger + a_1\right) + \hbar v_2\left(a_2^\dagger + a_2\right),$$

where $a_1$ is the bosonic annihilation operator for the optical mode around 1.8 eV that our COMSOL simulations reproduce, the one responsible for the coupling to the exciton (where $\sigma$ is the excitonic annihilation operator). Both are coupled through a Jaynes-Cummings term with strength $g$. On the other hand, $a_2$ is the annihilation operator for the exciton-decoupled, blue-detuned mode observed in the experiments. The electron static-like current couples to both optical modes, and not the TMD excitons, with relative strengths given by $v_i$. Assuming that the pumping is small, and in order to account for optical and excitonic radiative decay[15], we build an effective non-hermitian Hamiltonian of the form

$$H_{eff} = H_{pump} + \hbar\Omega_1 a_1^\dagger a_1 + \hbar\Omega_\sigma \sigma^\dagger \sigma + \hbar g\left(a_1^\dagger \sigma + a_1 \sigma^\dagger\right) + \hbar\Omega_2 a_2^\dagger a_2,$$

where $\Omega_i = \left(\omega_i - \frac{i\gamma_i}{2}\right)$ and, in the same fashion, $\Omega_\sigma = \left(\omega_\sigma - \frac{i\gamma_\sigma}{2}\right)$. Note that the $\gamma$'s are the radiative decay rates of the different elements of the system. We can diagonalize the terms that couple the optical mode $a_1$ and the exciton $\sigma$ above. These give rise to the polaritons observed experimentally. The polaritonic Hamiltonian is then given by $H_{pol} = \hbar\Omega_1 a_1^\dagger a_1 + \hbar\Omega_\sigma \sigma^\dagger \sigma + \hbar g(a_1^\dagger \sigma + a_1 \sigma^\dagger)$, and is diagonalized as $H_{pol}|N,\pm\rangle = \hbar\Omega_\pm(N)|N,\pm\rangle$, with $\Omega_\pm(N) = N\Omega_1 - \Delta \pm \sqrt{\Delta^2 + Ng^2}$, and

$$|N,\pm\rangle = \frac{g\sqrt{N}|n,g\rangle + (\Delta \mp \sqrt{\Delta^2 + Ng^2})|n-1,e\rangle}{\sqrt{Ng^2 + |\Delta \mp \sqrt{\Delta^2 + Ng^2}|^2}} \equiv A_\pm|n,g\rangle + B_\pm|n-1,e\rangle,$$

where $\Delta = (\Omega_1 - \Omega_\sigma)/2$, and $|g\rangle$ and $|e\rangle$ correspond, respectively, to the ground and excited state of the exciton, and $|n\rangle$ (where n = N) is the number of photons in mode $a_1$. Then, one can treat the pump term using first-order perturbation theory, and find the perturbed ground state of the system, which turns out to be

$$|0'\rangle \approx |0,g\rangle \otimes |0\rangle_2 - \nu_1 \left(\left[\frac{A_+^*}{\Omega_+^*}|1,+\rangle + \frac{A_-^*}{\Omega_-^*}|1,-\rangle\right] \otimes |0\rangle_2 + \frac{\nu_2}{\Omega_2^* \nu_1}|0,g\rangle \otimes |1\rangle_2\right).$$

Once this perturbed ground state is known, we can compute the power spectrum (under weak-pumping), given by

$$I(\omega) = Re\left(\lim_{T\to\infty} \frac{1}{2\pi T} \int_{-T/2}^{T/2} dt \int_{-\infty}^{\infty} d\tau \langle 0'|\xi^\dagger(t)\xi(t-\tau)|0'\rangle e^{-i\omega\tau}\right),$$

where $\xi = \vec{\mu_i} a_i$, with $\vec{\mu_i}$ being the effective dipole moment of the optical modes. We assume that detected emission mainly comes from the optical modes, and since the pumping is incoherent, we consider separately the intensity emitted by each optical mode (thus neglecting cross-correlation terms). The lineshapes for both modes are therefore

$$I_1(\omega) = \frac{1}{2\pi}|\nu_1\mu_1|^2 \left( \left|\frac{A_+^2}{\Omega_+}\right|^2 \frac{|\Gamma_+|}{(\omega-\omega_+)^2+\Gamma_+^2} + \left|\frac{A_-^2}{\Omega_-}\right|^2 \frac{|\Gamma_-|}{(\omega-\omega_-)^2+\Gamma_-^2} \right),$$

$$I_2(\omega) = \frac{1}{2\pi}|\nu_2\mu_2|^2 \left|\frac{1}{\Omega_2}\right|^2 \frac{|\Gamma_2|}{(\omega-\omega_2)^2+\Gamma_2^2},$$

where $\omega_i = Re(\Omega_i)$, and $\Gamma_i = Im(\Omega_i)$. Finally, the measured intensity is given by the projection of the far field amplitude over the polarizer, which can be calculated as

$$I_T(\omega) = I_1(\omega)\cos^2(\phi-\phi_1) + I_2(\omega)\cos^2(\phi-\phi_2),$$

where $\phi$ is the polarizer angle and $\phi_i$ is the orientation of the dipole moment of mode $i$. We set the parameters of the TMD exciton and first cavity mode to the values obtained from our numerical simulations: $\phi_1 = 45°$, $\omega_1 = 1.75$ eV, $\gamma_1 = 2\Gamma_1 = 90$ meV, $g = 35$ meV, $\omega_\sigma = 1.75$ eV, $\gamma_\sigma = 10$ meV. For the second cavity mode, we take $\omega_2 = 1.81$ eV, $\gamma_2 = 2\Gamma_2 = 140$ meV, and $\phi_2$ as described in the main text and in the caption of Fig. 4. The other free parameter is the ratio $f = |\nu_2\mu_2/\nu_1\mu_1|^2$, which weights how much more (or less) the two different cavity modes get excited in the near field by the tunneling current ($\nu_2/\nu_1$), and how relatively bright they are ($\mu_2/\mu_1$). These two mechanisms are indistinguishable from the far field. We know from experimental data that the $a_2$ mode is brighter, so we have set $f = 3$, with which we produce the results shown in the main text.

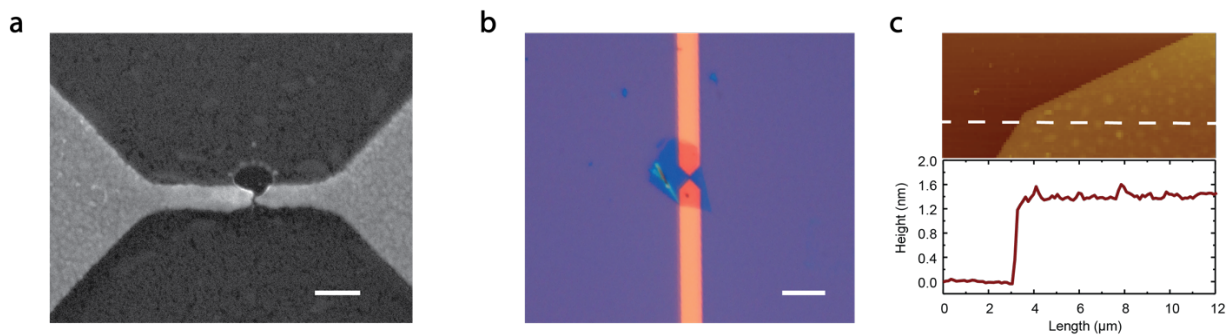

**Figure S1**. (a) SEM image showing the damaged TMD on top of the nano gap due to dielectric breakdown when the electrical bias across the gap exceeds 1.6 V. Scale bar in the figure is 200 nm. (b) Optical microscope image of a nanowire right after dry transfer of TMD. Scale bar in the image is 10 μm. (c) AFM scan image and height profile for freshly exfoliated monolayer $WSe_2$.

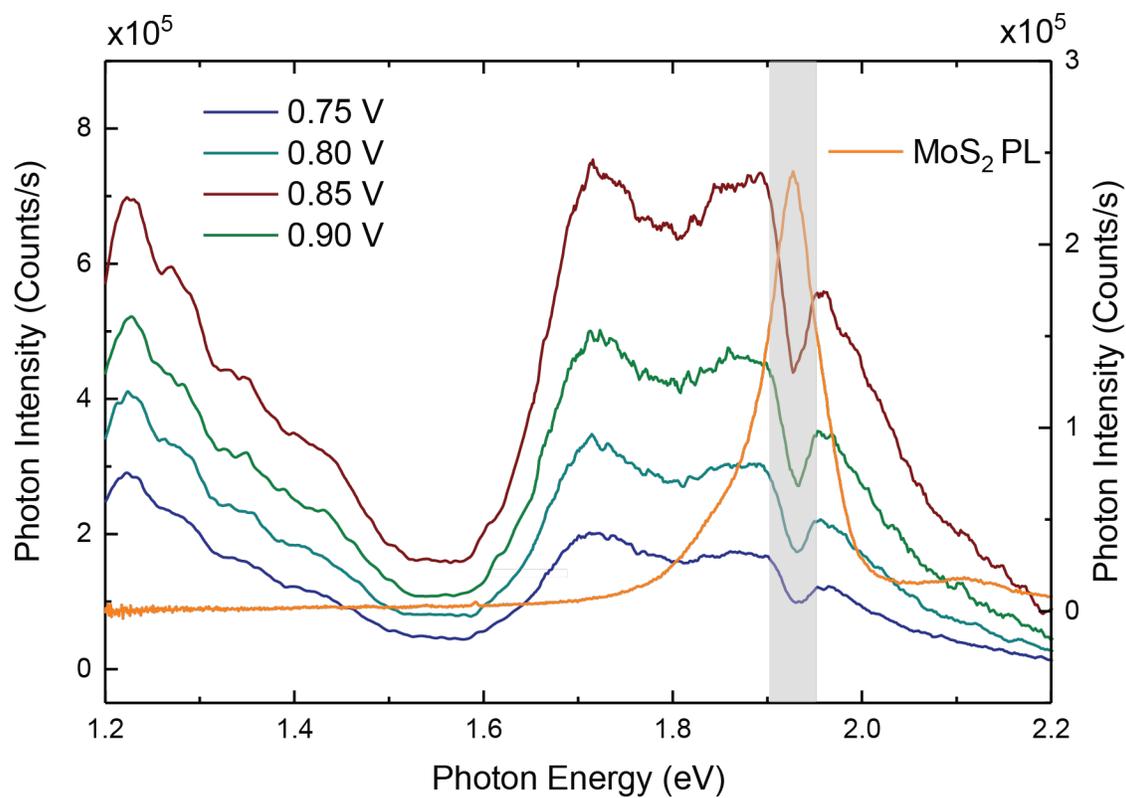

**Figure S2**. Spectral measurements of EL spectra with MoS$_2$ transferred on top of the nano gap. The spectra are overlaid with the PL spectra from bare MoS$_2$ for the same flake, highlighting the correspondence between the exciton peak and position of polaritonic dip.

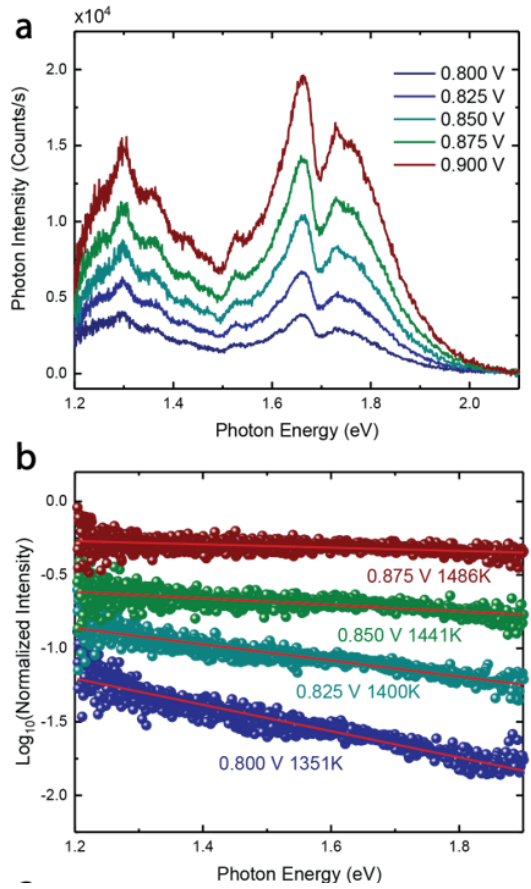

**Figure S3**. Normalization analysis to extract the optical radiative local density of states. (a) Full spectra of the EL spectra shown in Fig. 2a. (b) Logarithmic plot for normalized spectra versus photon energy, by dividing the measured spectrum at 0.800, 0.825, 0.850 and 0.875 V with reference to the spectrum at 0.900 V. The solid red lines correspond to linear fittings to a Boltzmann energy distribution $e^{-\hbar\omega/k_B T_{eff}}$. Extracted effective temperature $T_{eff}$ are labeled adjacent to each line.

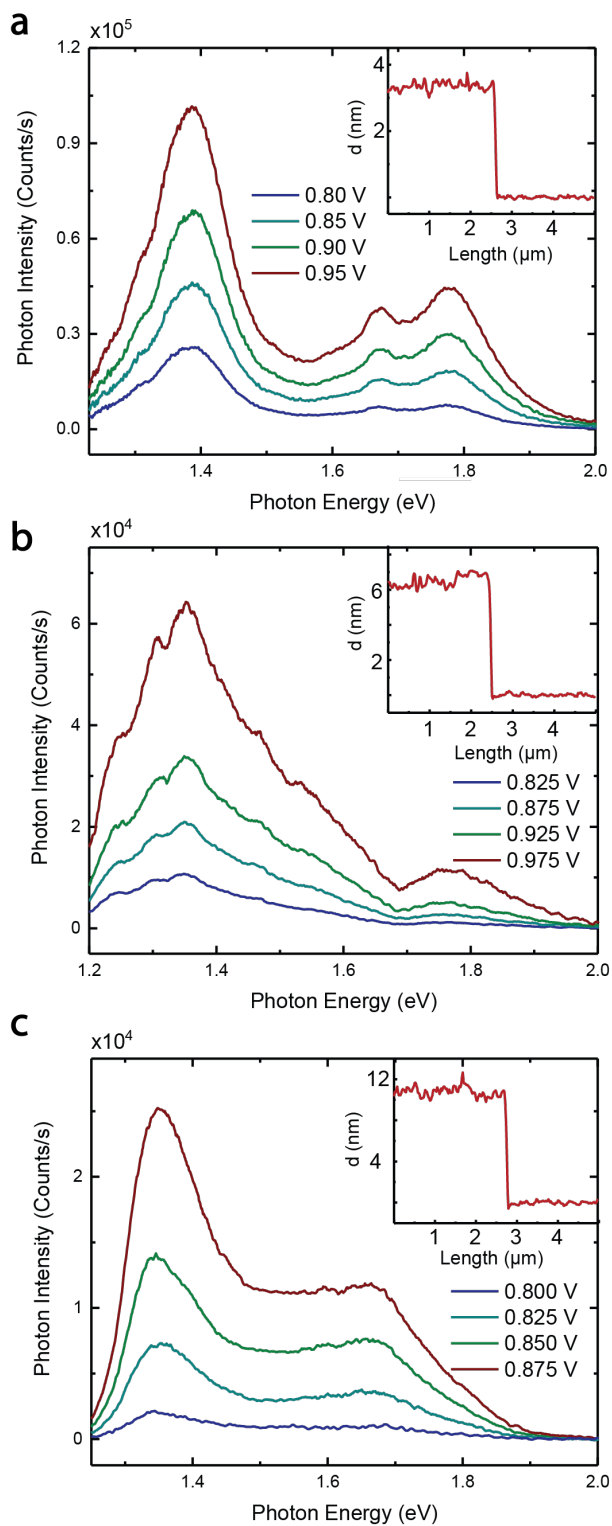

**Figure S4**. Spectral measurements of EL spectra for hybrid TMD-nanogap system with different thickness of $WSe_2$. The AFM profile of the thickness of the TMD is illustrated on the top right corner. (a) EL spectra from 0.80 V to 0.95 V for 2-layer $WSe_2$. (b) EL spectra from 0.825 V to 0.975 V for 4-layer $WSe_2$. (c) EL spectra from 0.800 V to 0.875 V for 7-layer $WSe_2$.

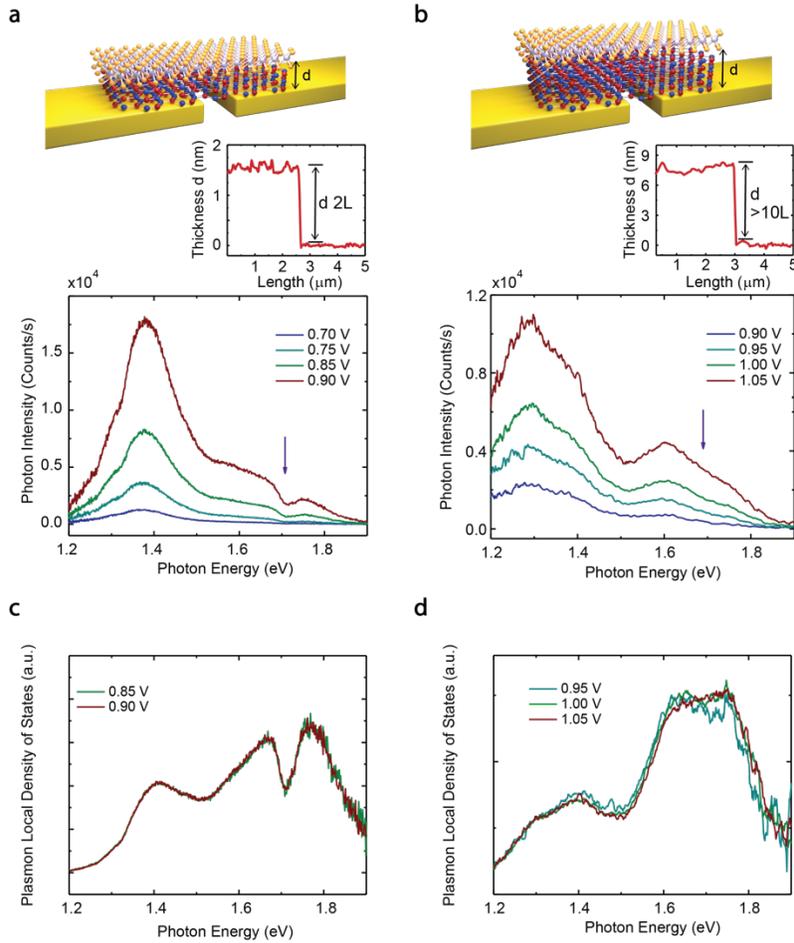

**Figure S5**. Spectral measurements of EL spectra for the hybrid TMD-on-gap structure spaced by hBN structures with different hBN thickness. (a) Measured EL spectra for the TMD-on-gap structure with bilayer hBN between the WSe$_2$ and the gold nanogap, from 0.7 V to 0.9 V (0.15 $G_0$ zero-bias conductance). The WSe$_2$ A exciton PL peak energy is indicated by the purple arrow. Height profile of the encapsulated hBN is shown in the middle inset. (b) Measured EL spectra at different biases from 0.9 V to 1.05 V with encapsulated more than 10 layers of hBN between the WSe2 and the gold nanogap ((0.27 $G_0$ zero-bias conductance). At this greater spacing between the TMD and the nanogap, the plasmon-exciton coupling is too small to lead to detectable spectral signatures. (c) Extracted plasmon local density of states for the measured spectra in (a). (d) Extracted plasmon local density of states for the measured spectra in (b).

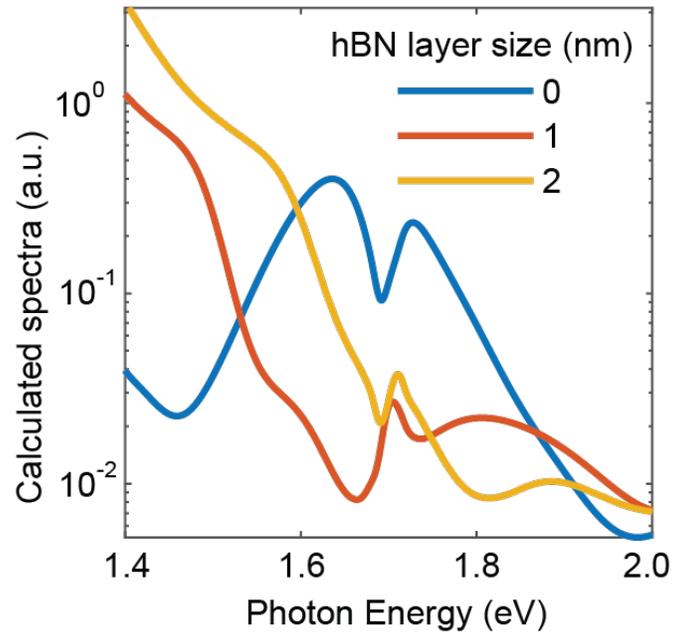

**Figure S6**. Numerically calculated spectra with various thickness of encapsulated hBN in between the TMD-gap system. It shows the case with no spacer, and for hBN layer sizes of 1 and 2 nm thicknesses.

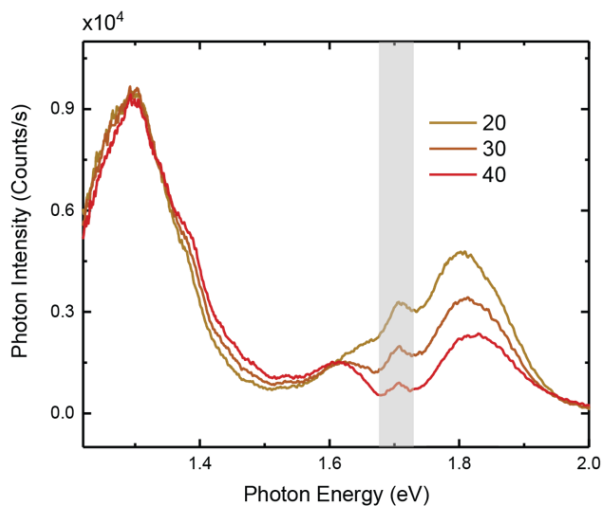

**Figure S7**. Measured polarization-selected EL spectra (polar contour plot of the spectra is shown in Fig. 4b) showing the feature at the exciton energy that can be attributed to the light scattered by WSe$_2$ excitons remaining uncoupled to the plasmon resonance. The legend labels the spectra by the angle in degrees of the detected polarization, where zero is along the nanowire direction.

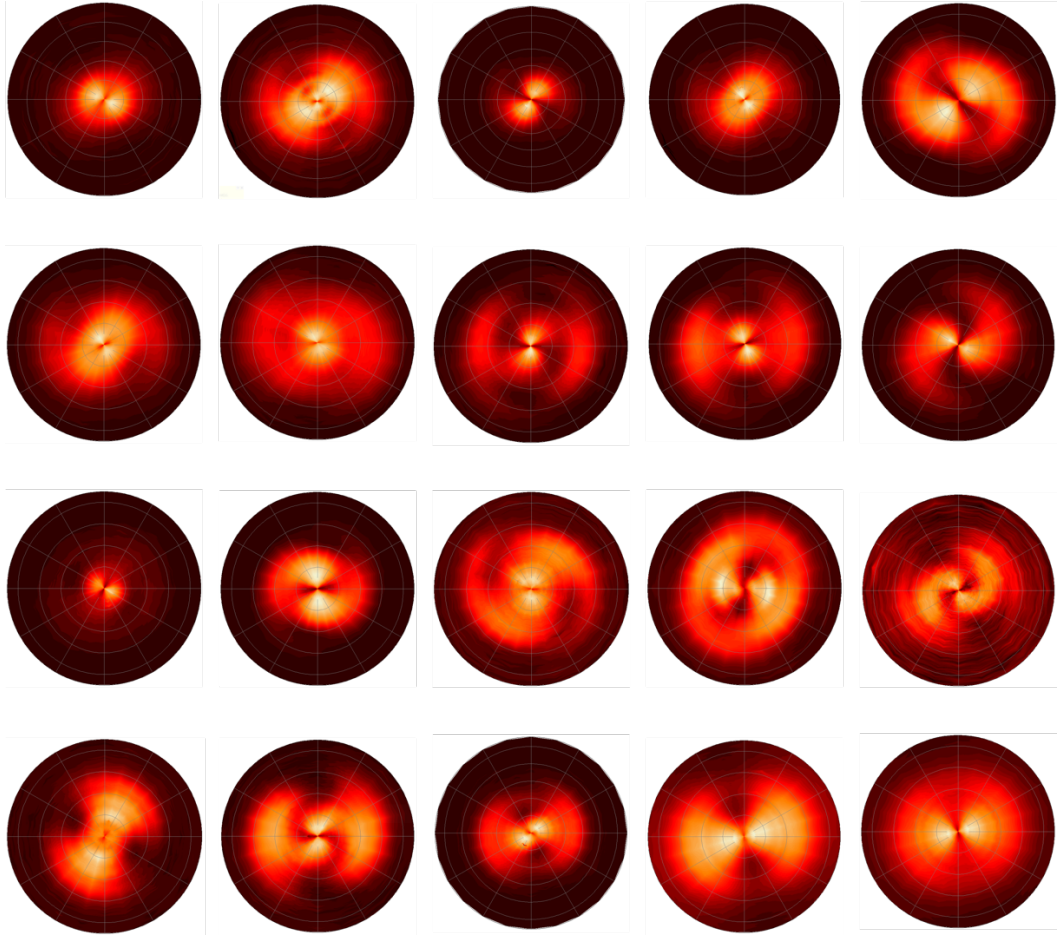

**Figure S8**. Summary of polar contour plots of polarization-selected light emission spectra from twenty representative bare-metal junctions. None of these show polariton-like peak splittings. The radial coordinate is the energy, ranging from 1.2eV to 2.1eV. The applied voltage ranges from 1.0V to 1.4V. For each plot, the color scale is normalized between the maximum photon counts (white) and the zero photon counts (black).

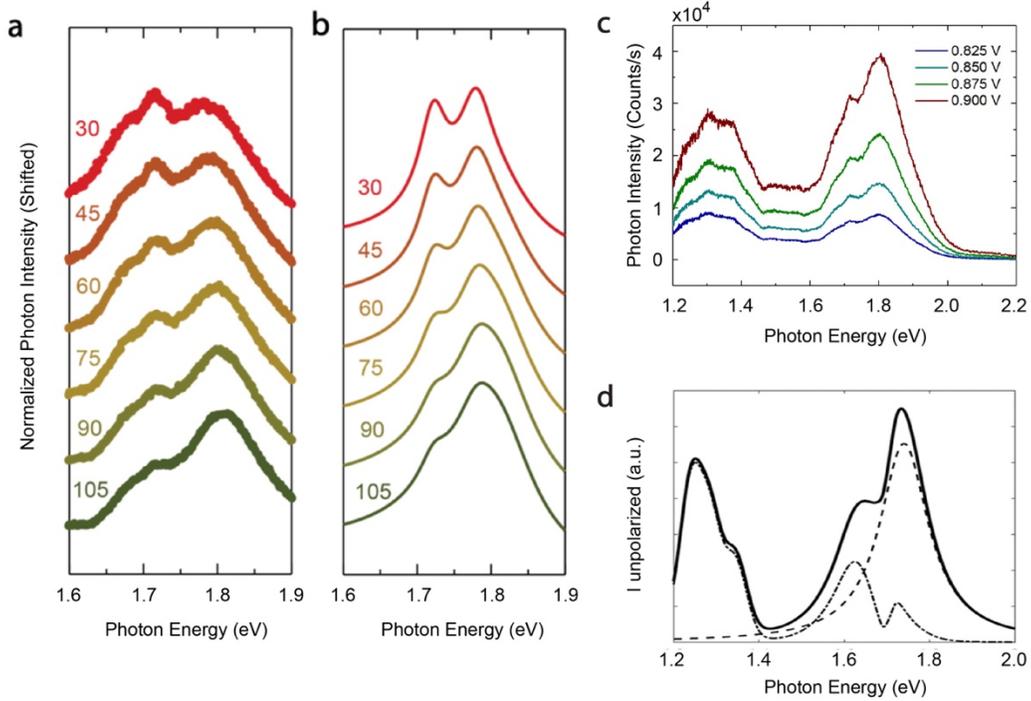

**Figure S9**. The two-plasmon model and polarization dependence. (a) Polarization-selected EL spectra at fixed bias 0.8V with a zero-bias junction conductance of 0.09 $G_0$ at different detection polarizations. These are fixed-angle cuts of the data shown in the full polarization plot in Fig. 4a. (b) Calculated polarized spectra based on the simplified model involving coupling between the exciton and one of two plasmon modes. (c) Non-polarization-selected spectra for the same device from biases of 0.825 V to 0.900 V. (d) Numerical results for the single plasmon-exciton coupling calculation (dot dash line) and the decoupled guided surface plasmon mode (dashed line), plotted together with the total unpolarized spectra (solid line).

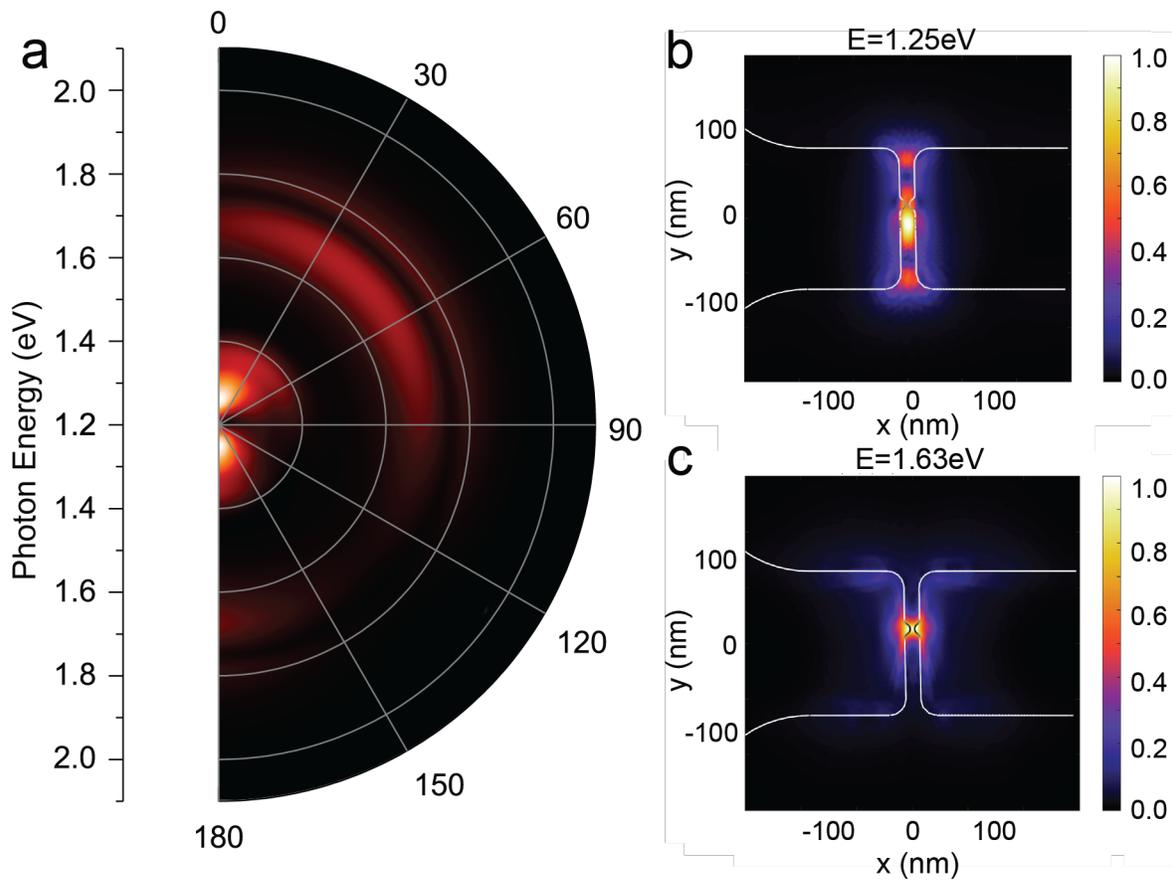

**Figure S10**. Modeling plasmons via finite-element methods. (a) Polar contour plot from the EM simulations successfully reproducing the low-energy (around 1.25 eV) LSP mode present in the devices, energetically very off resonance from the excitons, and the exciton coupled LSP mode (around 1.63 eV). (This is for the same device as Fig. 4a and Fig. S9, but omitting the non-exciton-coupled guided plasmon mode.) (b,c) Field distributions associated with the low energy LSP and 1.63 eV LSP (energetically resonant with the exciton), taken at $z$ set at the half-height of the TMD layer.

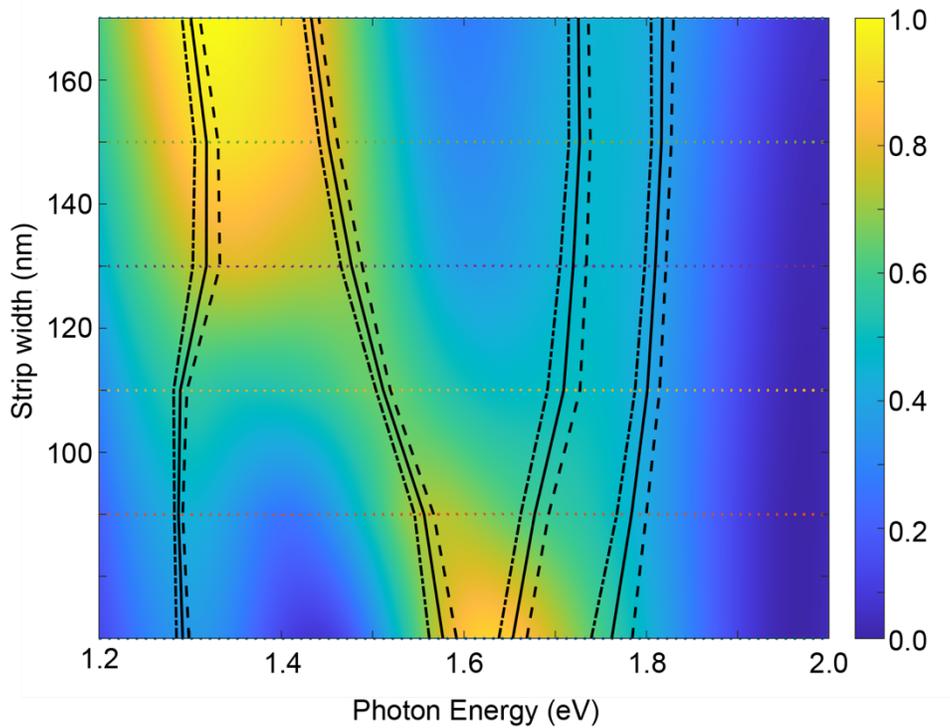

**Figure S11**. Scattered power by the nanowire upon plane-wave excitation as a function of nanowire width and photon energy. Solid black lines represent the central frequencies of a fit to a 4-Lorentzian function.

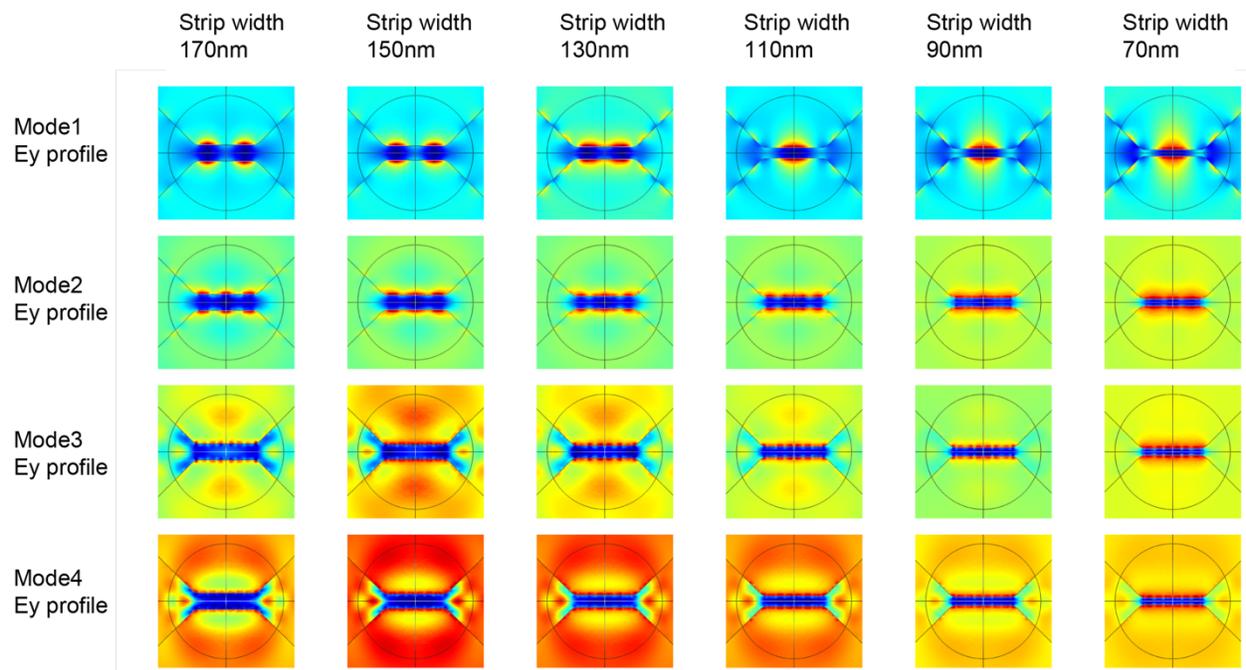

**Figure S12**. $y-$component of the electric field (normal to the wire axis) for the different modes identified in Figure S11 for different values of nanowire width. The modes are labelled in ascending photon energy order.

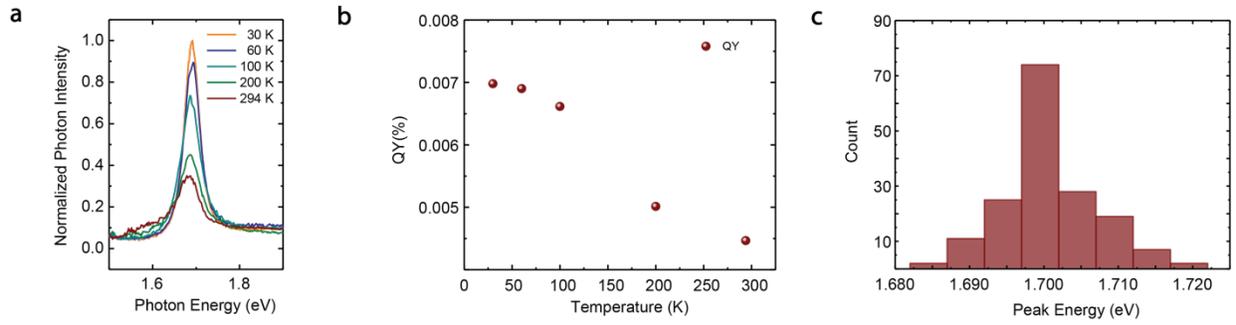

**Figure S13**. Temperature dependent and statistical analysis for PL map of a bilayer $WSe_2$. (a) Temperature dependent PL of $WSe_2$ from 30K to room temperature. The exciton peak intensity decreases as temperature rises. (b) Calculated PL quantum yield in (a) at different temperatures. (c) Statistical histogram of the exciton peak position when performing a PL map scan across a 10um x 10um area of the same TMD flake.

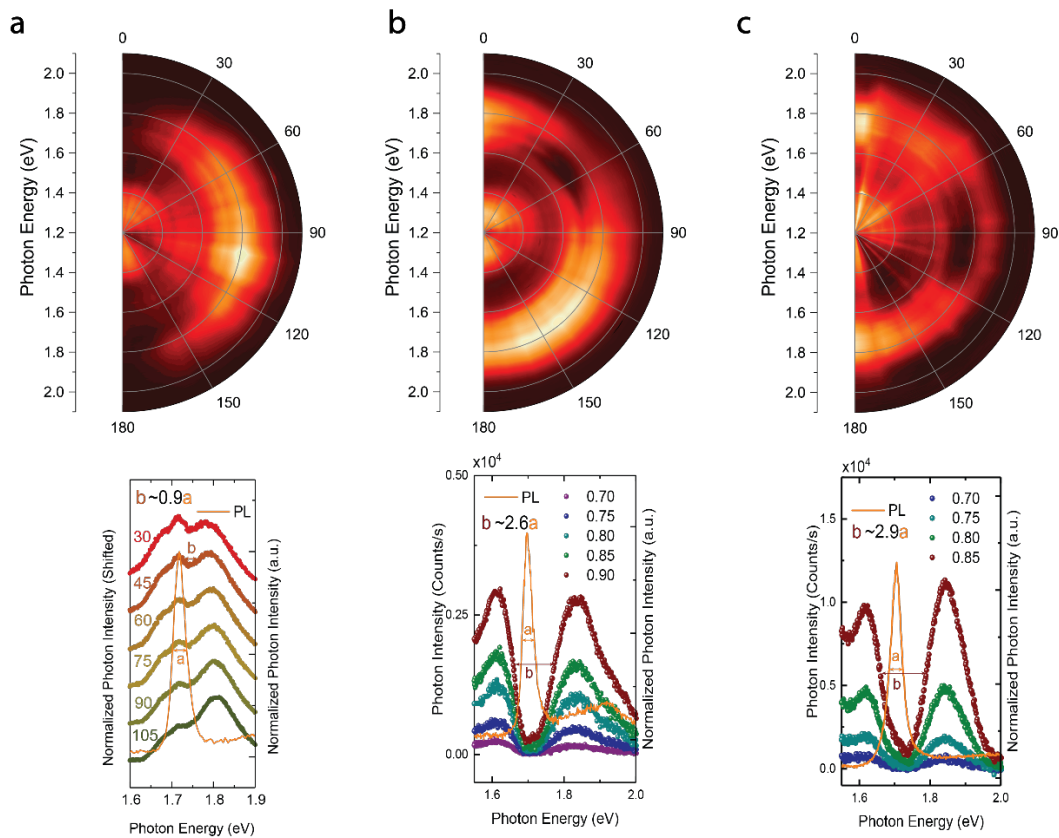

**Figure S14.** Unnormalized full contour plot of the experimental results in Fig. 4a, b and c and the linecut of the emission spectra overlayed with respective exciton PL spectra. (a) the upper panel of (a-c) shows the polar plot for the experimental measured spectra, while the lower panel of (a) shows the normalized and vertically shifted spectra at different polarization angles overlayed with the PL in orange. In (b-c) the lower panel shows the voltage dependent EL spectra at 45 degrees polarization for (b) and 105 degrees for (c) respectively. The orange line is the measured PL from each device. The FWHM linewidth of the exciton and mode splitting are denoted as '*a*' and '*b*', respectively, with the numerical relationship between these values displayed on the top left corner of the lower panels.

**Supporting Information References**